\documentclass[aps,amssymb, showkeys]{revtex4}
\usepackage{graphicx}
\usepackage{epstopdf}
\usepackage[pdftex, bookmarks, colorlinks=true, plainpages = false, citecolor = red, urlcolor = blue, filecolor = blue]{hyperref}
\usepackage{amsmath}

\newcommand{\be}{\begin{equation}}
\newcommand{\ee}{\end{equation}}
\newcommand{\bea}{\begin{eqnarray}}
\newcommand{\eea}{\end{eqnarray}}
 
\newcommand{\ba}[1]{\begin{array}{#1}}
\newcommand{\ea}{\end{array}}

\renewcommand{\b}{\beta}

\def\k{\kappa}
\renewcommand{\l}{\lambda}
\def\m{\mu}
\def\n{\nu}

\begin{document}

\title{Twist operator correlation functions in $O(n)$ loop models}

\author{Jacob J. H. Simmons}
\email{j.simmons1@physics.ox.ac.uk}
\affiliation{Rudolf Peierls Centre for Theoretical Physics, 1 Keble Road, Oxford OX1 3NP, UK}

\author{John Cardy}
\email{j.cardy1@physics.ox.ac.uk}
\affiliation{\mbox{Rudolf Peierls Centre for Theoretical Physics, 1 Keble Road, Oxford OX1 3NP, UK}\\
\emph{and}\\
All Souls College, Oxford}


%%%%%%%%%%%%%%%%%%%%
%                    ABSTRACT                        %
%%%%%%%%%%%%%%%%%%%%
\begin{abstract}
Using conformal field theoretic methods we calculate correlation functions of geometric observables in the loop representation of the $O(n)$ model at the critical point.  We focus on correlation functions containing twist operators, combining these with anchored loops, boundaries with SLE processes and with double SLE processes.

We focus further upon $n = 0$, representing self-avoiding loops, which corresponds to a logarithmic conformal field theory (LCFT) with $c=0$.   In this limit the twist operator plays the role of a $0-$weight indicator operator, which we verify by comparison with known examples.  Using the additional conditions imposed by the twist operator null-states, we derive a new explicit result for the probabilities that an SLE$_{8/3}$ wind in various ways about two points in the upper half plane, \emph{e.g.} that the SLE passes to the left of both points.

The collection of $c=0$\, logarithmic CFT operators that we use deriving the winding probabilities is novel, highlighting a potential incompatibility caused by the presence of two distinct logarithmic partners to the stress tensor within the theory.  We provide evidence that both partners do appear in the theory, one in the bulk and one on the boundary and that the incompatibility is resolved by restrictive bulk-boundary fusion rules.
\end{abstract}
\maketitle

%%%%%%%%%%%%%%%%%%%%
%                INTRODUCTION                   %
%%%%%%%%%%%%%%%%%%%%
\section{Introduction}

It is well established that many two dimensional statistical physics systems, including the $O(n)$ and percolation models which we focus on for this paper, can be mapped to an equivalent loop representation \cite{Nienhuis83}. Much of the recent interest in these loop representations of statistical systems has been motivated by the success of Schramm-Loewner evolution (SLE) in describing loops emanating from boundaries \cite{Schramm00,RohdeSchramm05}.  Progress has also been made in quantifying the effects of the fluctuations due to background loops, dubbed the `loop-soup', via conformal loop ensembles (CLE) \cite{Werner08}.

Alternately the loop model can be mapped to a height model via the Coulomb gas, which renormalizes onto a conformal field theory (CFT) when taken to the continuum limit.  This allows us to study these loop models using the powerful methods of CFT \cite{BPZ84, Cardy84}.   However, the unitary minimal model CFTs that are most familiar from the study of statistic mechanics are insufficient to describe the full behavior that can be observed in the loop models.  We are forced to consider the extended Kac table, which implies the existence of logarithmic modules in the CFT.  Logarithmic (L)CFTs have been studied extensively \cite{Gurarie93,EberleFlohr06}, but little progress has been made in completing the association to general loop models.

In \cite{Cardy00} operators were identified in the Couloumb gas that when applied in sets of two, change the weight of all loops that separate the insertion points.   When the new weight is chosen to be the negative of the original weight we have the twist operators, first studied in \cite{GamsaCardy06}.  Inserting a pair of twist operators modifies the partition function with a positive or negative weight for each configuration depending on the parity of the number of loops crossing a defect line between the two points.  When background loops are suppressed (such as in the $n=0$ limit of the $O(n)$ model, which is identified with self avoiding walks and loops) the correlation functions can be used either directly or through a small $n$ expansion to determine the winding properties of loops around the twist points.  

Regularization issues arise associated to small loops and are addressed in \cite{GamsaCardy06}  by inserting additional twist operators and isolating those solutions where loops separate pairs of twist operators.  The distance between the twist operators sets a scale preventing the profusion of small loops.

In this paper we we fix the scale by other means: In the bulk we fix a scale for our loops using $2-$leg operators that ensure that the loops pass through two given points.  The result is Eq.\ (\ref{SALExpr}) the probability within the ensemble of self avoiding loops anchored at $z_3$ and $z_4$ that the loop separates $z_1$ from $z_2$.   

In regions with boundaries we use boundary $N-$leg operators to set our scale. We begin with a pair of $1-$leg or SLE operators, and include only one twist operator, which marks parity against the boundary, and recover Schramm's result for the left crossing SLE probability \cite{Schramm01} at $\k=8/3$.    Next we include a result for two self avoiding walks anchored at common points using a pair of boundary $2-$leg operators along with a twist operator.  We see that at $\k = 8/3$ this is equivalent to a result first reported in \cite{GamsaCardy05} for double self avoiding walks.

We then extend Schramm's result for the self-avoiding walk by including a second twist operator which allows us to determine the weights of SLE$_{8/3}$ conditioned to wind about two points in any of four possible ways.  For example, given an SLE$_{8/3}$ process in the upper-half plane from $0$ to infinity the probability of a \emph{double} left passage with respect to the two points $r_A\, \mathrm{exp}(i v_A)$ and $r_B\, \mathrm{exp}(i v_B)$ is 
\bea \label{PLL}
\frac{1}{4}\left( (1+\cos(v_A))(1+\cos(v_B))+\sin(v_A)\sin(v_B)\left(1-\sigma\,{}_2F_1\left(1,\frac{4}{3};\frac{5}{3}\bigg|1-\sigma \right)\right)\right)\; ,
\eea
where we have defined
\be
\sigma=\frac{r_A{}^2-2r_A r_B \cos(v_A-v_B)+r_B{}^2}{r_A{}^2-2r_A r_B \cos(v_A+v_B)+r_B{}^2}\; .
\ee
This result is the half-plane version of (\ref{WABStrip}); equations (\ref{WABStrip}-\ref{W0Strip}) convey all four of the winding probabilities in either upper-half plane or infinite strip coordinates.

This analysis also provides an additional example of a physically meaningful $c=0$ LCFT.  Recently a great deal of attention has been paid to the boundary (B)CFT of critical percolation \cite{SimmonsKlebanZiffJPA07, MathieuRidout07, Kytola08}, which is an example of a $c=0$ LCFT generated by a zero weight module with second order null state. This CFT possesses a transparent interpretation in terms of the equivalent SLE$_6$ arc representations.  In the $O(n=0)$ model the chiral components of twist operator belong to the same LCFT module as the percolation SLE operator.  Thus the $O(n=0)$ twist operators give us a second physical picture with which to probe this LCFT.  

In addition, we find an example of how the holomorphic and anti-holomorphic sectors combine to give physical correlation functions in the bulk, and we note that the leading terms of fusions associated to bulk logarithmic modules tend to have non-zero spin.  We identify one such fusion that yields a weight one, spin one, leading fusion product for all values of $n$, indicating the presence of a non-Kac zero weight staggered logarithmic module in all corresponding CFTs.

Also, in \cite{GurarieLudwig00,MathieuRidout07} the point was made that at $c=0$ the logarithmic theories generated by the two modules with second order null states, $\mathfrak{M}_{1,2}$ and $\mathfrak{M}_{2,1}$, are incompatible in the sense that the two point function between logarithmic operators from the two theories cannot be defined.  This has lead to speculation that the two CFTs cannot both be subsets of in the same physical theory.   However, our calculation of (\ref{PLL}) requires solving a correlation function containing bulk twist operators ($\phi_{2,1}$) and boundary SLE operators($\phi_{1,2}$) and we find a set of solutions satisfying all applicable second order null state relation.  This refutes the conjecture that it is inherently unphysical to combine these modules. We include a brief analysis of our solutions, leading to the conjecture that the two LCFT generated by $\mathfrak{M}_{1,2}$ and $\mathfrak{M}_{2,1}$ may both be sub-theories of the same BCFT so long as one appears in the bulk theory, the other appears in the boundary theory, and the bulk-boundary fusions all act via the identity module.

In section \ref{O(n)} we calculate several twist operator correlation functions in the $O(n)$ model for general $n$.  In section \ref{AnchSAL}-\ref{2SLETwist} we specialize these results to the self avoiding loop ensemble.  In \ref{NewResult} we derive a new SLE$_{8/3}$ result by solving a six point chiral correlation function. In section \ref{LCFT} we discuss the implications of this chiral correlation function for the $c=0$ LCFT.  Our conclusions are in section \ref{Conc}.

%%%%%%%%%%%%%%%%%%%%
%           O(n) LOOPS W/ TWISTS           %
%%%%%%%%%%%%%%%%%%%%
\section{The $O(n)$ loops with twist operators. \label{O(n)}}

We begin this section with a brief description of the loop representation of the $O(n)$ model and and quick summary of the steps required associate it to a meaningful CFT.  For a more complete treatment we refer the interested reader to the pertinent references.

The standard loop representation of the $O(n)$ model \cite{Nienhuis83} begins with $n-$component spins $\boldsymbol{s}(r_i)$ with squared norm $n$ on each site of a lattice $\Omega$ governed by the partition function
\be
Z_{\Omega} = \mathrm{Tr} \prod_{\langle i j \rangle}\left( 1+x\, \boldsymbol{s}(r_i) \cdot \boldsymbol{s}(r_j)\right)\; .
\ee
We expand the product and associate a graph on $\Omega$ to each term by including the bond between  $r_i$ and $r_j$ if the $x\, \boldsymbol{s}(r_i) \cdot \boldsymbol{s}(r_j)$ factor appears in the factor and excluding the bond if it does not.  Now the trace per site of an odd number of spin components is zero, so the only graphs that contribute are those composed entirely of closed loops.  The form of the resulting loop partition function is particularly simple if we choose the honeycomb lattice $HC$ for $\Omega$.  In this case the loops can visit each site a maximum of one time and the per site trace, $\mathrm{Tr} s_a(r_i)s_b(r_i) = \delta_{a b}$, means that each loop earns a net weight $n$.  Combining this with the weight $x$ per occupied bond yields the partition function
\be
Z_{HC}=\sum_{\Lambda} n^\mathcal{N} x^L\; ,
\ee
where the sum is over all loop configurations $\Lambda$ on the honeycomb lattice, $\mathcal{N}$ the number of loops, and $L$ the total length of the loops in each configuration.  For small values of $x$, long loops are repressed and the model flows to the vacuum under renormalization.  For large values of $x$ long loops carry more weight and the system flows to a fixed point of densely packed loops under renormalization.  The boundary between these two regimes is $x_c=(2+\sqrt{2-n})^{-1/2}$, for which the system is critical and flows to the dilute fixed point \cite{Nienhuis83}.  

We can map the loop model to the Coulomb gas by replacing the sum over configurations, with a sum over configurations of \emph{directed} loops with a complex local weight that recreates the factors related to the non-local observable $\mathcal{N}$.   To recreate these factors we associate a weight $\exp(i \chi \theta / 2)$ to each vertex containing a loop, where $\theta$ is the angle the loop turns though while traversing the vertex.  Each loop mush close on itself thereby picking up a total weight of $\exp(\pm i \chi \pi)$ depending on whether it is directed clockwise or counter-clockwise.  Taking the trace over the directions in each directed loop configuration is equivalent to an undirected loop configuration with weight $2 \cos(\chi\, \pi)$ per loop, so that by careful selection of $\chi$ we recover the $O(n)$ model weights.

The directed loops are equivalent to level lines of a height variable on the dual of our lattice if we insist that the height variable increases (decreases) by $\pi$ whenever we cross a loop pointing to the left (right).  The power of the Coulomb gas formalism lies in the assumption that this height model flows into a Gaussian free field under renormalization, which allows us to make precise calculations in the continuum limit of these models using field theory techniques.  

There has been a great deal of success in using operators in the height model to glean information about the associated loop model and to derive relations between the model and the CFT describing its continuum limit.  We emphasize that our description of the mappings between these models is by no means complete, as there are a variety of subtleties that we simplify or omit completely.

It can be argued that the two non-trivial fixed points of the $O(n)$ loop model correspond to SLE/CLE$_{\k}$s with
\be
n=2 \cos\left(\frac{(\k -4)\pi}{\k}\right)\;,\quad \left\{ \begin{array}{l}(2<\k<4)\quad\mathrm{dilute}\\( 4<\k)\quad\mathrm{dense}\end{array} \right.\; .
\ee
%Notice that we've used $4/\k$ for the Coulomb gas parameter $g$ as has become standard in the SLE literature.  
Alternately the correspondence with the Coulomb gas implies that these loop models are described by CFTs with central charge and conformal weights given by
\be
c=\frac{(6-\k)(3\k-8)}{2\k}\; ,\quad\mathrm{and}\quad h_{r,s}=\frac{(\k\, r-4\, s)^2-(\k-4)^2}{16 \k}\; .
\ee

In this article we focus primarily on the dilute regime with $2<\k<4$ relating to the critical $O(n)$ model.  In \cite{GamsaCardy06}  twist operators were identified that change the weight of all loops that separate  two twist operators so that the new weight associated to these loops is $-n$.   The partition function becomes
 \be
Z=\sum_{\Lambda} (-1)^{N_s} n^N x^L\; ,
\ee
where $N_s$ is the number of loops separating the twist operators.  In \cite{GamsaCardy06} it was shown that the twist operator was a Kac operator with weight
\be
h_{\mathrm{twist}} = \bar h_{\mathrm{twist}} = h_{2,1}=\frac{3 \kappa -8}{16}\; .
\ee
In this paper we utilize the second order null-state descendant of this Kac operator to derive differential equations in twist operator correlation functions.

%%%%%%%%%%%%%%%%%%%%
%         ANCH. LOOPS W/ TWIST           %
%%%%%%%%%%%%%%%%%%%%
\subsection{Twist operators and anchored $O(n)$ loops in the bulk \label{O(n)BulkAnch}}
In order to avoid the complications that small loops cause to twist operators, we are interested in anchoring our bulk loops with $2-$leg operators to establish a scale for our loops.  In \cite{SaleurDuplantier87}, the exponent for these operators are calculated in the Coulomb gas formulation of the loop model.  In the Coulomb gas the $2-$leg operators are equivalent to the insertion of a vortex and and anit-vortex, where the (anti-)vortex has $2$ directed lines flowing into (out of) the point.  Because of this, no path can have both of its ends connected to a single $2-$leg operator and furthermore, the paths associated to these vortices have weight $1$ not $n$.  Using the Coulomb gas vertex association it can be shown \cite{SaleurDuplantier87,Cardy05SLE} that these operators have weight
\be
h_{2-\mathrm{leg}}=\bar h_{2-\mathrm{leg}}=h_{0, 1}=\frac{8-\k}{16}\; ,
\ee
where we use the Kac weight convention to give the weight of these operators even though the indices are not positive integers.  This means that the $2-$leg operators do not have null state descendants, nor related differential equations, for general $\k$.

The correlation function we calculate includes one pair of $2-$leg operators and one pair of twist operators.  As per the usual CFT approach, we focus on the holomorphic sector,
\be
\langle \phi_{2,1}(z_1) \phi_{2,1}(z_2)\phi_{0,1}(z_3)\phi_{0,1}(z_3) \rangle = (z_2-z_1)^{-2 h_{2,1}} (z_4-z_3)^{-2 h_{0,1}}F\left( \frac{(z_2-z_1)(z_4-z_3)}{(z_3-z_1)(z_4-z_2)}\right)\; ,
\ee
 and will later sew this together with its anti-holomorphic counterpart \cite{BPZ84}.

The null state $\left( 2 (1+2 h_{2,1})L_{-2}-3 L_{-1}{}^2\right)|\phi_{1,2}\rangle$ implies that the differential operator
\be
3 \partial_{z_1}{}^2-2(1+2 h_{2,1})\left( \frac{h_{0,1}}{(z_4-z_1)^2}-\frac{\partial_{z_4}}{z_4-z_1}+\frac{h_{0,1}}{(z_3-z_1)^2}-\frac{\partial_{z_3}}{z_3-z_1}+\frac{h_{2,1}}{(z_2-z_1)^2}-\frac{\partial_{z_2}}{z_2-z_1} \right)\; ,
\ee
annihilates the correlation function \cite{BPZ84}.  If we apply this differential operator to the correlation function and take the standard limit $\{ z_1,z_2,z_3,z_4\} \to \{0,x,1,\infty\}$ then we recover the following condition for $F(x)$;
\be\label{DE1}
0=F''(x)+\frac{(8-2\k)-(8-\k) x}{4 x(1-x)}F'(x)-\frac{\k (8-\k)}{64(1-x)^2} F(x)\; .
\ee

Solving (\ref{DE1}) we find the $x \approx 0$ conformal blocks
\bea \label{DE11}
F_{1,1}(x) &=&(1-x)^{-\k/8}{}_2F_1\left(-\frac{\k}{4},1-\frac{\k}{4};2-\frac{\k}{2}; x  \right)\quad \mathrm{and}\\ \label{DE12}
F_{3,1}(x) &=&x^{\k/2-1}(1-x)^{-\k/8}{}_2F_1\left(\frac{\k}{4}-1,\frac{\k}{4};\frac{\k}{2}; x  \right)\; ,
\eea
where the subscript corresponds to the corresponding fusion as $z_2 \to z_1$.  The fusion products in $\phi_{2,1} \times \phi_{2,1} = \mathbf{1}+\phi_{3,1}$ are the identity and the leading order energy density operator.

In contrast, there is an ambiguity in the conformal blocks as $z_2 \to z_3$.  The fusion, $\phi_{2,1} \times \phi_{0,1} = \phi_{1,1}+\phi_{-1,1}$, results in operators with weights $0$ and $1$ for all values of kappa.  These weights differ by an integer, which is indicative of a staggered logarithmic module.    In particular this module must have three key operators: $\Phi_0$, the primary highest weight operator; $\partial \Phi_0$, which is orthogonal to all the other descendants of $\Phi_0$ but couples to; $\Phi_1$, the logarithmic partner to $\partial \Phi_0$, and the highest weight generator of the module.   This structure is similar to the module $\mathfrak{I}_{1,4}$ discussed in \cite{MathieuRidout07}, but appears for all values of kappa, not just those with $c=0$.  Speculatively, this module may be related to the zero weight indicator operator used to gauge the position of a point relative to a given set of paths, e.g. in Schramm's derivation of the SLE left passage probability \cite{Schramm01}.

The nature of the staggered module is such that we cannot fix a unique expression for the corresponding conformal block $G_{1,1}(1-x)$ becasue we are free to arbitrarily modify our choice by adding a multiple of $G_{-1,1}(1-x)$.

We write an expression for the $x \approx 1$ blocks that includes this ambiguity.  We do this by applying the logarithmic series expansion for the hypergeometric function, found for example in \cite{AbSt},  to (\ref{DE11}) and (\ref{DE12}) and combining these blocks to have a leading term of unity 
\bea \label{DE21}
G_{1,1}(1-x)&=&A \frac{\Gamma\left(1-\frac{\k}{4}\right) \Gamma\left(2-\frac{\k}{4}\right)}{\Gamma\left(2-\frac{\k}{2}\right)} F_{1,1}(x)+(1-A)\frac{\Gamma\left(1+\frac{\k}{4}\right) \Gamma\left(\frac{\k}{4}\right)}{\Gamma\left(\frac{\k}{2}\right)} F_{3,1}(x)\quad \mathrm{and}\\ \label{DE22}
G_{-1,1}(1-x)&=&(1-x)^{1-\k/8}{}_2F_1\left(1-\frac{\k}{4},-\frac{\k}{4};2;1-x\right)\; .
\eea
We will discuss the implications of this ambiguity in greater detail once we've obtained physical solutions for the correlation function.

We now introduce a convenient normalization for the conformal blocks:
\bea
\mathcal{F}_{1,1}(x) &=&F_{1,1}(x)\; ,\\
\mathcal{F}_{3,1}(x) &=&\frac{\Gamma\left(\frac{4+\k}{4}\right)\Gamma\left(\frac{\k}{4}\right)\Gamma\left(\frac{4-\k}{2}\right)}{\Gamma\left(\frac{\k}{2}\right)\Gamma\left(\frac{4-\k}{4}\right)\Gamma\left(\frac{8-\k}{4}\right)}F_{3,1}(x)\; ,\\
\mathcal{G}_{1,1}(1-x)&=&\frac{\Gamma\left(\frac{4-\k}{2}\right)}{\Gamma\left(\frac{4-\k}{4}\right)\Gamma\left(\frac{8-\k}{4}\right)}G_{1,1}(1-x)\; ,\quad \mathrm{and}\\
\mathcal{G}_{-1,1}(1-x)&=&\frac{\Gamma\left(\frac{4+\k}{4}\right)\Gamma\left(\frac{\k}{4}\right)}{\Gamma\left(\frac{\k-2}{2}\right)}G_{-1,1}(1-x)\; .
\eea
We normalize the leading term of $\mathcal{F}_{1,1}(x)$ to one in order to recover 
\be
\langle \phi_{2,1}(z_1) \phi_{2,1}(z_2)\phi_{0,1}(z_2)\phi_{0,1}(z_2) \rangle  \approx \langle \phi_{2,1}(z_1) \phi_{2,1}(z_2)\rangle\, \langle\phi_{0,1}(z_2)\phi_{0,1}(z_2) \rangle 
\ee
in the limit $|z_1-z_2|$ and $|z_3-z_4| \ll |z_1-z_3|$.  The other blocks are normalized to simplify the crossing relations, which become
\bea \nonumber
\mathcal{F}_{1,1}(x)&=&\mathcal{G}_{1,1}(1-x)+(1-A)\, \mathcal{G}_{-1,1}(1-x)\; ,\\ \label{Xsym}
\mathcal{F}_{3,1}(x)&=&\mathcal{G}_{1,1}(1-x)-A\, \mathcal{G}_{-1,1}(1-x)\; ,\\ \nonumber 
\mathcal{G}_{1,1}(1-x)&=&A\, \mathcal{F}_{1,1}(x)+(1-A)\, \mathcal{F}_{3,1}(x)\;,\quad \mathrm{and}\\ \nonumber
\mathcal{G}_{-1,1}(1-x)&=&\mathcal{F}_{1,1}(x)-\mathcal{F}_{3,1}(x)\; .
\eea

We need to study the monodromy of our blocks as we move $x$ around the points $0$ and $1$, in order to construct a single-vaued physical solution.
Moving clockwise around zero, the $x \approx 0$ blocks transform as
\bea \label{0mon}
\mathcal{F}_{1,1}(x)&\stackrel{0 \circlearrowright}{\longrightarrow}& \mathcal{F}_{1,1}(x) \\ \nonumber
\mathcal{F}_{3,1}(x)& \stackrel{0 \circlearrowright}{\longrightarrow}& e^{-\k \pi i} \mathcal{F}_{3,1}(x)\; .
\eea
%The crossing symmetry relations then imply
%\bea
%\mathcal{G}_{1,1}(1-x)&\stackrel{0 \circlearrowright}{\longrightarrow}& (1+A(1-e^{-\k \pi i}))\mathcal{G}_{1,1}(1-x)+A(1-A)(1-e^{-\k \pi i})\mathcal{G}_{-1,1}(1-x)\\ \nonumber
%\mathcal{G}_{-1,1}(1-x)&\stackrel{0 \circlearrowright}{\longrightarrow}&(1-e^{-\k \pi i})\mathcal{G}_{1,1}(1-x)+(1-A(1-e^{-\k \pi i}))\mathcal{G}_{-1,1}(1-x)
%\eea
%based on the transformations of the $x \approx 0$ conformal blocks.

The effect of moving around $1$ is more complicated due to the logarithm in the expansion of $\mathcal{G}_{1,1}(1-x)$ around $x=1$.  To determine the effect of these logarithmic terms we isolate them from the remaining algebraic terms by applying the logarithmic series expansion to (\ref{DE21}).  We are careful to maintain our modified normalizations and after some simplification of the gamma functions we find that
\be
\mathcal{G}_{1,1}(1-x) = (1-x)^{-\k/8}S(1-x)+\frac{1}{2 \pi} \tan\left(\frac{\pi \k}{4}\right) \log(1-x) \mathcal{G}_{-1,1}(1-x)\; ,
\ee
where $S(1-x)$ is the regular power series contribution, which is regular about the point $x=1$.  On the other hand, taking $x$ clockwise around $1$ the logarithm transforms as
\be
\log(1-x) \stackrel{1 \circlearrowright}{\longrightarrow} \log(1-x)-2 \pi i\; .
\ee
Thus moving clockwise around $1$ we find that
\bea 
\mathcal{G}_{1,1}(1-x)&\stackrel{1 \circlearrowright}{\longrightarrow}& e^{\k \pi i/4}\mathcal{G}_{1,1}(1-x)-i\, e^{\k \pi i/4} \tan\left(\frac{\pi \k}{4}\right)  \mathcal{G}_{-1,1}(1-x)\; ,\quad \mathrm{and} \\  \nonumber
\mathcal{G}_{-1,1}(1-x)&\stackrel{1 \circlearrowright}{\longrightarrow}&e^{\k \pi i/4}\mathcal{G}_{-1,1}(1-x)\; ,
\eea
which can be combined with (\ref{Xsym}) yielding
\bea \label{1mon}
\mathcal{F}_{1,1}(x)&\stackrel{1 \circlearrowright}{\longrightarrow}&\sec\left(\frac{\k \pi}{4}\right) \mathcal{F}_{1,1}(x)+i\, e^{\k \pi i/4} \tan\left(\frac{\k \pi}{4}\right) \mathcal{F}_{3,1}(x)\\ \nonumber
\mathcal{F}_{3,1}(x)& \stackrel{1 \circlearrowright}{\longrightarrow}&-i\, e^{\k \pi i/4}  \tan\left(\frac{\k \pi}{4}\right) \mathcal{F}_{1,1}(x)+e^{\k \pi i/2}\sec\left(\frac{\k \pi}{4}\right)\mathcal{F}_{3,1}(x)\; .
\eea

This completes our analysis of the chiral solution space.  Now we need to sew together holomorphic and anti-holomorphic conformal blocks into a single valued physical solution.  Based on (\ref{0mon}) we see that the physical solution must be of the form $\mathcal{F}_{1,1}(x)\mathcal{F}_{1,1}(\bar x)+C\mathcal{F}_{3,1}(x)\mathcal{F}_{3,1}(\bar x)$\; .  If we combine this with the condition that the correlation function remain unchanged when we move $x$ around $1$, then we can uniquely determine the physical solution
\be
|\mathcal{F}_{1,1}(x)|^2-|\mathcal{F}_{3,1}(x)|^2 = \mathcal{G}_{1,1}(1-x)\mathcal{G}_{-1,1}(1-\bar x)+\mathcal{G}_{-1,1}(1-x)\mathcal{G}_{1,1}(1-\bar x)+(1-2A)|\mathcal{G}_{-1,1}(1-x)|^2\; .
\ee
The negative coefficient serves as a reminder that breaking the Kac symmetry so that $\phi_{2,1} \ne \phi_{1,1}$ and including the operator $\phi_{0,1}(z,\bar z)$ from the boundary of the Kac table leads to a non-unitary theory.
 
Restoring all relevant factors and the implicit dependence on the four coordinates we have
\bea \nonumber
&&\langle \phi_{2,1}(z_1,\bar z_1) \phi_{2,1}(z_2, \bar z_2)\phi_{0,1}(z_3, \bar z_3)\phi_{0,1}(z_4, \bar z_4) \rangle =\\ \label{Expr1}
&&\phantom{killingspace}\left|\frac{z_{21}}{z_{43}} \right|^2 \left|\frac{z_{43}z_{31}z_{42}}{z_{21}{}^3 z_{32} z_{41}} \right|^{\k/4}\left[ \left| {}_2F_1\left(-\frac{\k}{4},\frac{4-\k}{4};\frac{4-\k}{2};\frac{z_{21}z_{43}}{z_{31}z_{42}} \right) \right|^2 \right. \\ \nonumber
&&\phantom{killingspace}\left. -\frac{\k^2 \Gamma\left(\frac{\k}{4} \right)^4\Gamma\left(\frac{4-\k}{2} \right)^2}{(4-\k)^2\Gamma\left(\frac{\k}{2} \right)^2\Gamma\left(\frac{4-\k}{4} \right)^4} \left|\frac{z_{21}z_{43}}{z_{31}z_{42}} \right|^{\k -2} \left| {}_2F_1\left(\frac{\k-4}{4},\frac{\k}{4};\frac{\k}{2};\frac{z_{21}z_{43}}{z_{31}z_{42}} \right) \right|^2 \right]\; ,
\eea
where we've used the shorthand $z_{ij}:=z_i-z_j$.

%%%%%%%%%%%%%%%%%%%%
%          B'DRY PATHS W/ TWIST           %
%%%%%%%%%%%%%%%%%%%%
\subsection{Boundary $O(n)$ paths and twist operators}

In the previous sections we found an expression for a simple bulk correlation function using $2-$leg operators and twist operators.  In this section we derive analogous correlation functions in regions with a boundary, anchoring paths to the boundary instead of loops in the bulk.   In addition to the twist operators we use the boundary $N-$leg operators.  These can be identified in the Coulomb gas as operators that change the boundary conditions by $N$ steps within some neighborhood of their insertion.  This leads to the identification with the boundary operator
\be
h_{N-\mathrm{leg}}=h_{1,N +1}=\frac{N\, (4+2 N -\k)}{2 \k}\; .
\ee

In particular we emphasize that the correlation functions we calculate in this section differ from the boundary correlation functions in \cite{GamsaCardy06} in that we only insert a single twist operator into our correlation functions.  The effect of the twist operator in this case is to indicate the parity of loops separating the insertion point and boundary.  If we have loops attached to the boundary the twist operator we may have an unimportant ambiguity in the overall sign depending on which part of the boundary we choose to measure our parity from.

We first examine the correlation function
\be
\langle \phi_{2,1}(z,\bar z)\phi_{1,2}(x_1) \phi_{1,2}(x_2) \rangle 
\ee
which uses a $1-$leg boundary operator to encode an SLE process in the presence of a twist operator.

The correlation function takes the form
\be
(z-\bar z)^{-2h_{2,1}}(x_2-x_1)^{-2h_{1,2}} F\left( \frac{(z-\bar z)(x_2-x_1)}{(z-x_1)(x_2-\bar z)}\right)\; ,
\ee
and the null states of $\phi_{2,1}$ and $\phi_{1,2}$ imply that it is annihilated by
\bea
3 \partial_{z}{}^2-2(1+2 h_{2,1})\left( \frac{h_{2,1}}{(\bar z-z)^2}-\frac{\partial_{\bar z}}{\bar z-z}+\frac{h_{1,2}}{(x_1-z)^2}-\frac{\partial_{x_1}}{x_1-z}+\frac{h_{1,2}}{(x_2-z)^2}-\frac{\partial_{x_2}}{x_2-z} \right)&& \mathrm{and}\\
3 \partial_{x_1}{}^2-2(1+2 h_{1,2})\left( \frac{h_{2,1}}{(z-x_1)^2}-\frac{\partial_ z}{z-x_1}+\frac{h_{2,1}}{(\bar z-x_2)^2}-\frac{\partial_{\bar z}}{\bar z-x_2}+\frac{h_{1,2}}{(x_2-x_1)^2}-\frac{\partial_{x_2}}{x_2-x_1} \right)&&\mathrm{respectively}.
\eea

Applying these differential operators to the correlation function and letting $\{ \bar z, z,x_2,x_1\} \to \{0,x,1,\infty\}$, we find two differential equations governing $F(x)$;
\bea \label{DE2}
0&=&F''(x)+\frac{2(4-\k)-(8-\k) x}{4 x(1-x)}F'(x)-\frac{(6-\k)}{8(1-x)^2} F(x)\quad \mathrm{and}\\
0&=&F''(x)-\frac{2(4-\k)-(4-2 \k) x}{\k\, x(1-x)}F'(x)-\frac{(3 \k-8)}{4 \k (1-x)^2} F(x)\; .
\eea
The only common solution to both of these differential equations is
\be
F(x)=\frac{2-x}{2 \sqrt{1-x}}\; .
\ee

This correlation function takes on a simple form in the strip $\mathbb{S}=\{ s=u+v i | t \in \mathbb{R},0<u<\pi  \}$, mapping the two anchoring points to $\pm \infty$ using
\be
s(z) = \log\left( \frac{z-x_1}{z-x_2} \right)\; ,
\ee
so that the cross ratio is
\be
x=\frac{(z-\bar z)(x_2-x_1)}{(z-x_1)(x_2-\bar z)}=1-\frac{(\bar z-x_1)(x_2-z)}{(z-x_1)(x_2-\bar z)} = 1-e^{2 i v}\; .
\ee
Thus, in the strip the correlation function can be written
\be \label{SLEeq}
\langle \phi_{2,1}(s,\bar s)\phi_{1,2}(-\infty) \phi_{1,2}(\infty) \rangle_{\mathbb{S}}  = \langle \phi_{2,1}(s,\bar s)\rangle_{\mathbb{S}} \langle\phi_{1,2}(-\infty) \phi_{1,2}(\infty) \rangle_{\mathbb{S}} \cos(v)\; ,
\ee
which is independent of our choice of $\k$.

We also include the correlation function
\be
\langle \phi_{2,1}(z,\bar z)\phi_{1,3}(x_1) \phi_{1,3}(x_2) \rangle 
\ee
describing a combination of a twist operator and a $2-$leg boundary operator.  The two paths generated by the $2-$leg operators can be thought of as a double SLE: Two SLE process starting from a common point and driven towards a common point, but conditioned not to meet.  

Again the correlation function is of the form
\be
(z-\bar z)^{-2h_{2,1}}(x_2-x_1)^{-2h_{1,3}} F\left( \frac{(z-\bar z)(x_2-x_1)}{(z-x_1)(x_2-\bar z)}\right)\; ,
\ee
except that now the null states of $\phi_{2,1}$ and $\phi_{1,3}$ imply that it is annihilated by
\bea
3 \partial_{z}{}^2-2(1+2 h_{2,1})\left( \frac{h_{2,1}}{(\bar z-z)^2}-\frac{\partial_{\bar z}}{\bar z-z}+\frac{h_{1,2}}{(x_1-z)^2}-\frac{\partial_{x_1}}{x_1-z}+\frac{h_{1,2}}{(x_2-z)^2}-\frac{\partial_{x_2}}{x_2-z} \right)&& \mathrm{and}\\
\partial_{x_1}{}^3-2(1+h_{1,3})\left( \frac{h_{2,1}\partial_{x_1}}{(z-x_1)^2}-\frac{\partial_ z\partial_{x_1}}{z-x_1}+\frac{h_{2,1}\partial_{x_1}}{(\bar z-x_2)^2}-\frac{\partial_{\bar z}\partial_{x_1}}{\bar z-x_2}+\frac{h_{1,3}\partial_{x_1}}{(x_2-x_1)^2}-\frac{\partial_{x_2}\partial_{x_1}}{x_2-x_1} \right)&&\\ \nonumber
-h_{1,3}(1+h_{1,3})\left( \frac{2 h_{2,1}}{(z-x_1)^3}-\frac{\partial_ z}{(z-x_1)^2}+\frac{2 h_{2,1}}{(\bar z-x_2)^3}-\frac{\partial_{\bar z}}{(\bar z-x_2)^2}+\frac{h_{1,3}}{(x_2-x_1)^3}-\frac{\partial_{x_2}}{(x_2-x_1)^2} \right)&&\mathrm{respectively}.
\eea

The corresponding differential equations are
\bea
0&=&F''(x) + \frac{2(4-\k)-(8-\k)x}{4x(1-x)} F'(x) -\frac{8-\k}{4(1-x)^2}F(x)\quad \mathrm{and}\\ \nonumber
0&=&F'''(x)-\frac{2(16-3 \k+(3\k-8)x)}{\k x(1-x)}F''(x)\\
&&+ \frac{6(8-\k)(4-\k)+4(6-\k)(3\k-8)x-(8-\k)(3\k-8)x^2}{\k^2 x^2 (1-x)^2} F'(x) +\frac{(3 \k-8)(8-\k)(2-x))}{\k^2 x (1-x)^3}F(x)
\eea
which only have one common solution:
\be
F(x)=1+\frac{8-\k}{6-\k}\, \frac{x^2}{4\,(1-x)}.
\ee
If we write this in terms of the strip geometry we find
\be \label{2xSLEEq}
\langle \phi_{2,1}(s,\bar s)\phi_{1,3}(-\infty) \phi_{1,3}(\infty) \rangle_{\mathbb{S}}  = \langle \phi_{2,1}(s,\bar s)\rangle_{\mathbb{S}} \langle\phi_{1,3}(-\infty) \phi_{1,3}(\infty) \rangle_{\mathbb{S}} \left(1-\frac{8-\k}{6-\k}  \sin^2(v) \right)\; .
\ee

%%%%%%%%%%%%%%%%%%%%
%          ANCH. SAL W/ TWIST                %
%%%%%%%%%%%%%%%%%%%%
\section{The anchored self avoiding loop with twist operators \label{AnchSAL}}

We have an expression for the correlator for any  $n \in (-2,2)$, but for a minute let's let $n =0$, or correspondingly $\k = 8/3$, where $h_{\mathrm{twist}}=0$ and $h_{2-\mathrm{leg}}=1/3$.  In this limit the $O(n)$ loop model corresponds to the self avoiding walk or self avoiding loop.  Strictly speaking, at $n=0$ all loops are suppressed and the only contribution to the partition function is the empty set so that $Z=1$.  In \cite{GamsaCardy06} this was circumnavigated by taking a small $n$ expansion and keeping the first order term, which was equivalent to conditioning the configurations on the existence of at least one loop.  While we need to do the same thing in spirit,  our job is made trivial by the inclusion of the $2-$leg operators, which obviously requires the existence of at least one loop and guarantees that that loop exists.  So, at $n=0$ we are restricted to those configurations with an self avoiding loop (SAL) connecting $z_3$ and $z_4$.  The total weight of these allowed configurations is given by the two point function
\be
Z = \langle \phi_{0,1}(z_3,\bar z_3) \phi_{0,1}(z_4,\bar z_4) \rangle = |z_{34}|^{-4/3}\; .
\ee

\begin{figure}[htb] %  figure placement: here, top, bottom, or page
   \centering
   \includegraphics[width=3in]{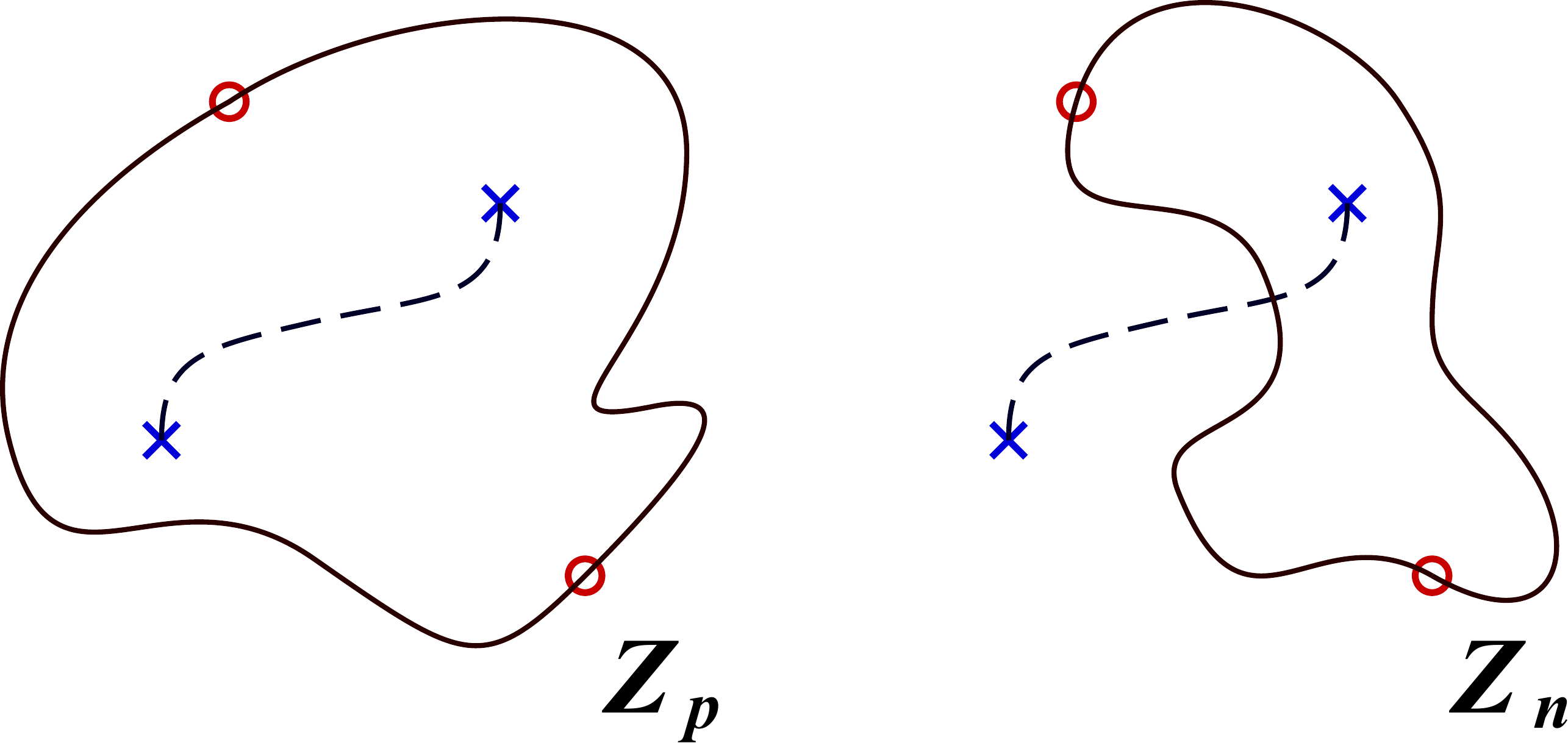} 
   \caption{The two distinct weights for a pair of $2-$leg operators (red circles) a pair of  twist operators (blue crosses) at $n=0$.}
   \label{conf1}
\end{figure}

With the inclusion of the twist operators we can further decompose $Z$ into two separate weights based on how the anchored SAL interacts with the twist defect.  The two relevant possibilities are illustrated in Fig.\ \ref{conf1}.
%FIGURE
The weight of configurations in which the anchored loop segregates the two twist defects is denoted by $Z_n$.  The weight of configurations where the loop fails to separate the twist defects is $Z_p$.  The index refers to whether the weight of the contribution is positive or negative with respect to the twist defect.

Now in terms of these weights it should be apparent that
\bea
Z&=&Z_p+Z_n \quad \mathrm{and}\\
Z_{\mathrm{twist}}&=&Z_p-Z_n\; 
\eea
with the twist partition function given by letting $\k =8/3$ in (\ref{Expr1})
\bea \label{Expr2}
Z_{\mathrm{twist}} &=&\left|\frac{z_{31}z_{42}}{z_{43}{}^2 z_{32} z_{41}} \right|^{2/3}\left[ \left| {}_2F_1\left(-\frac{2}{3},\frac{1}{3};\frac{2}{3};\frac{z_{21}z_{43}}{z_{31}z_{42}} \right) \right|^2 \right. \\ \nonumber
&&\left. -\frac{4 \Gamma\left(\frac{2}{3} \right)^6}{\Gamma\left(\frac{4}{3} \right)^2\Gamma\left(\frac{1}{3} \right)^4} \left|\frac{z_{21}z_{43}}{z_{31}z_{42}} \right|^{2/3} \left| {}_2F_1\left(-\frac{1}{3},\frac{2}{3};\frac{4}{3};\frac{z_{21}z_{43}}{z_{31}z_{42}} \right) \right|^2 \right]\; .
\eea
This allows us to determine the probability that a SAL separates $z_1$ and $z_2$ in the upper half plane, given that the loop passes through $z_3$ and $z_4$,
\bea \nonumber
P_n(z_1,z_2;z_3,z_4) &=& \frac{Z_n}{Z} =\frac{Z-Z_{\mathrm{twist}}}{2 Z}\\ \label{SALExpr}
&=&\frac{1}{2}\left[1-\left|\frac{z_{31}z_{42}}{z_{32} z_{41}} \right|^{2/3}\left| {}_2F_1\left(-\frac{2}{3},\frac{1}{3};\frac{2}{3};\frac{z_{21}z_{43}}{z_{31}z_{42}} \right) \right|^2 \right.\\ \nonumber
&& \left. +\frac{4 \Gamma\left(\frac{2}{3} \right)^6}{\Gamma\left(\frac{4}{3} \right)^2\Gamma\left(\frac{1}{3} \right)^4} \left|\frac{z_{21}z_{43}}{z_{32}z_{41}} \right|^{2/3} \left| {}_2F_1\left(-\frac{1}{3},\frac{2}{3};\frac{4}{3};\frac{z_{21}z_{43}}{z_{31}z_{42}} \right) \right|^2 \right]\; .
\eea
%FIGURE
\begin{figure}[htbp] %  figure placement: here, top, bottom, or page
   \centering
   \includegraphics[width=6in]{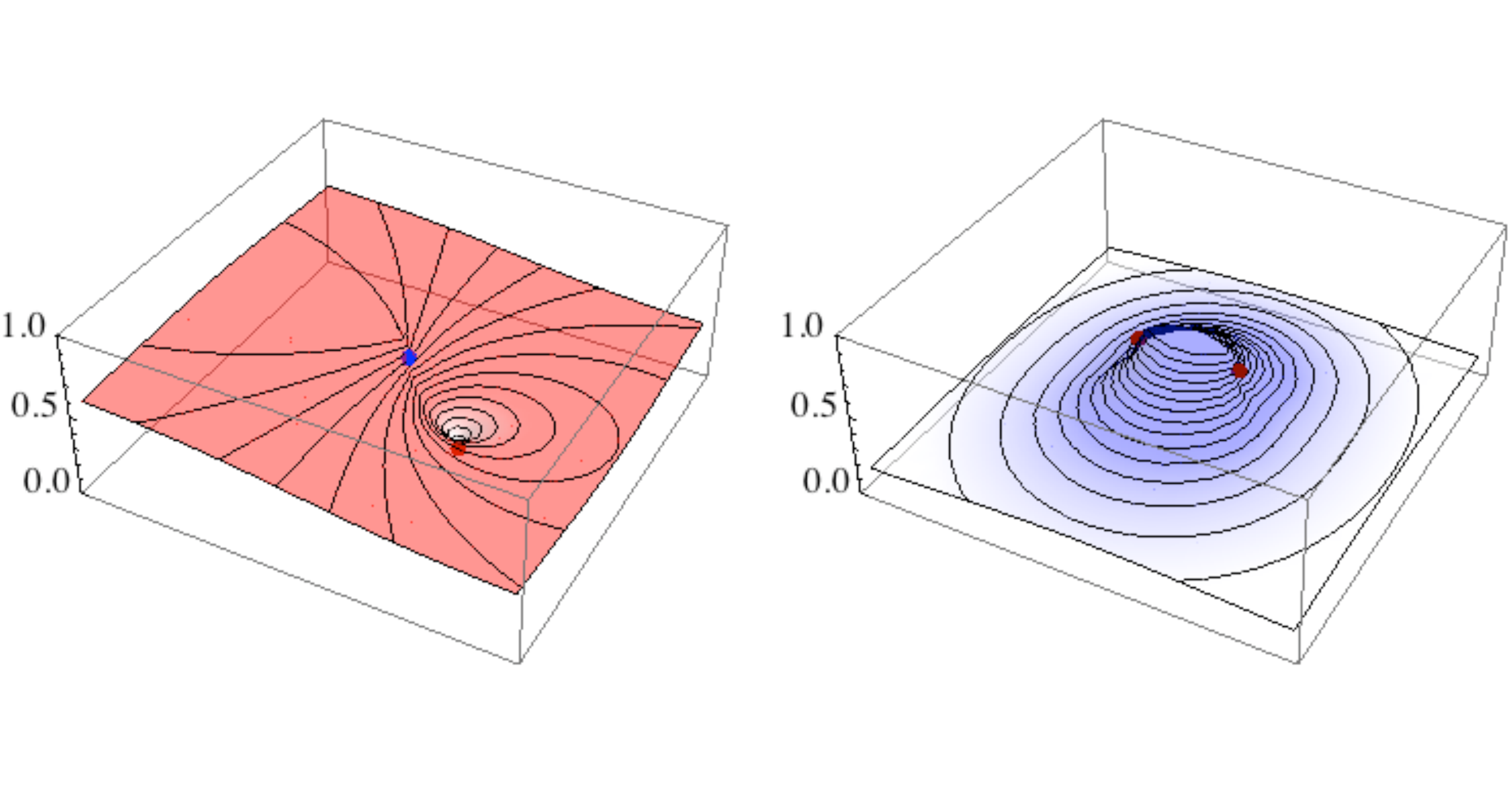} 
   \caption{On the left is a  plot of $P_n(0,\infty; 1, \bullet)$, on the right is a plot of $P_n(\bullet ,\infty; 1, -1)$.  Red dots indicate the positions of the $2-$leg operators.  The Blue dot indicates the position of the twist defect which sits away from infinity.}
   \label{geomcorr}
\end{figure}
As the cross-ratio $x=z_{21}z_{43}/z_{31}z_{42}$ goes to $0$ the twist defects and loop are well separated and the probability that the loop segregates the twist defects also goes to zero.   Maximum value for $P_n$ occurs when cross ratio goes to $2$.  Becasue of global conformal symmetry we are free to fix the position of three of our operators.  If we place the two $2-$leg operatorsa at $\pm1$ and one of the twist operators at infinity, then placing the second twist operator at the origin should maximize the probability of sitting inside the anchored loop, this corresponds to $x=2$.
The value of $P_n$ for $x=2$ is
\be
P_n^{\textrm{max}}=\frac{1}{2}+\frac{9\, \Gamma(5/6)^6}{4 \pi^3} \approx 0.6501\ldots\; .
\ee

As a final comment for $n=0$ we note that the weight $Z_p$ is equal to the related correlation function with the twists replaced by magnetization operators.  Pairs of magnetization operators behave in a fashion akin to pairs of twist operators; they changes the weight of the loops that separate the two points of insertion.  But whereas twist operators take $n \to -n$ the magnetization operators take $n \to 0$.  Since there can be no loop (or cluster hull) which separates them from each other, the magnetization operators measure configurations where the two points reside in the same cluster, which is exactly $P_p$.  The magnetization operator is $\phi_{1/2,0}$ \cite{Nienhuis83} and for $n=0$ it has conformal weight $h_{1/2,0}=0=h_{2,1}$.  The fact that these weights are equal is a necessary condition for their correlation functions to have common solutions.
\bea \label{MagExpr}
\langle \phi_{1/2,0}(z_1,\bar z_1) \phi_{1/2,0}(z_2, \bar z_2)\phi_{0,1}(z_3, \bar z_3)\phi_{0,1}(z_4, \bar z_4) \rangle &=&\frac{1}{2}+\frac{1}{2}\left|\frac{1}{1-x}\right|^{2/3}\left| {}_2F_1\left(-\frac{2}{3},\frac{1}{3};\frac{2}{3}; x \right) \right|^2\\ \nonumber
&&-\frac{2 \Gamma\left(\frac{2}{3} \right)^6}{\Gamma\left(\frac{4}{3} \right)^2\Gamma\left(\frac{1}{3} \right)^4} \left|\frac{x}{1-x} \right|^{2/3} \left| {}_2F_1\left(-\frac{1}{3},\frac{2}{3};\frac{4}{3}; x \right) \right|^2\; .
\eea

%%%%%%%%%%%%%%%%%%%%
%                    SLE W/ TWIST                    %
%%%%%%%%%%%%%%%%%%%%
\section{SLE$_{8/3}$ with a twist operator \label{SLETwist}}

\begin{figure}[htbp] %  figure placement: here, top, bottom, or page
   \centering
   \includegraphics[width=3in]{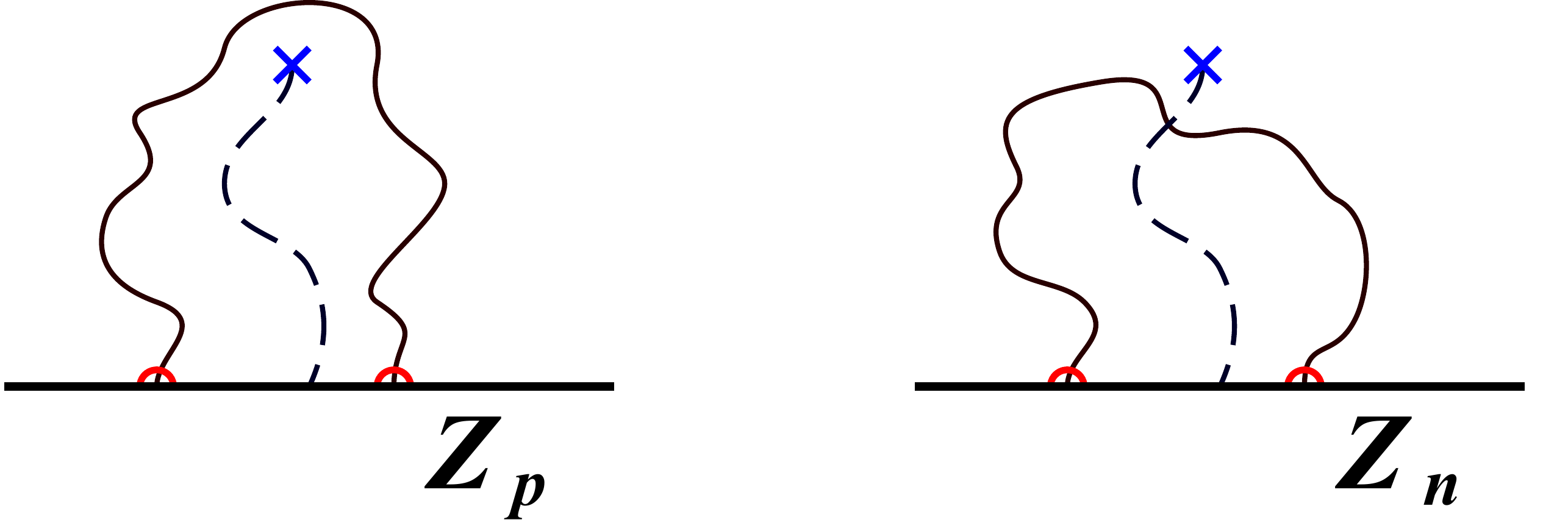} 
   \caption{The two different relevant contributions of the SLE (Red circles) process with respect to the defect point (Blue Crosses).}
   \label{BCFT1}
\end{figure}

We return to the correlation function
\be
\langle \phi_{2,1}(z,\bar z)\phi_{1,2}(x_1) \phi_{1,2}(x_2) \rangle\; ,
\ee
restricting ourselves to $n=0$ and therefore to an SLE$_{8/3}$ process.
We condition our correlation functions upon the existence of an SLE path from $x_1$ to $x_2$ so that the partition function is
\be
Z =Z_p+Z_n =  \langle \phi_{1,2}(x_1) \phi_{1,2}(x_2) \rangle = (x_2-x_2)^{-5/4}\;, 
\ee
dividing the ensemble of configurations that contribute to this partition function into two parts as in Fig.\ \ref{BCFT1}.  

Because the twist operators are zero weight operators we may rewrite the result in the strip (\ref{SLEeq}) as
\be
\frac{\langle \phi_{2,1}(s,\bar s)\phi_{1,2}(-\infty) \phi_{1,2}(\infty) \rangle_{\mathbb{S}}}{\langle\phi_{1,2}(-\infty) \phi_{1,2}(\infty) \rangle_{\mathbb{S}}} =  \frac{Z_p-Z_n}{Z} =  \cos(v)\; ,
\ee
highlighting the result as the twist operator correlation function conditioned on the SLE$_{8/3}$.   The probability that the SLE separates the twist operator from the bottom edge of the strip is
\be \label{SLEsingletwist}
P_p(v)=\frac{Z_p}{Z} = \frac{1+\cos{v}}{2}=\cos^2(v/2)\; .
\ee
As we expect, this is equivalent to Schramm's Formula for the for the left crossing probability in the strip with $\k = 8/3$ \cite{Schramm01}.

%%%%%%%%%%%%%%%%%%%%
%             DOUB. SLE W/ TWIST             %
%%%%%%%%%%%%%%%%%%%%
\section{Double SLE$_{8/3}$ with a twist operator \label{2SLETwist}}

\begin{figure}[htbp] %  figure placement: here, top, bottom, or page
   \centering
   \includegraphics[width=4.5in]{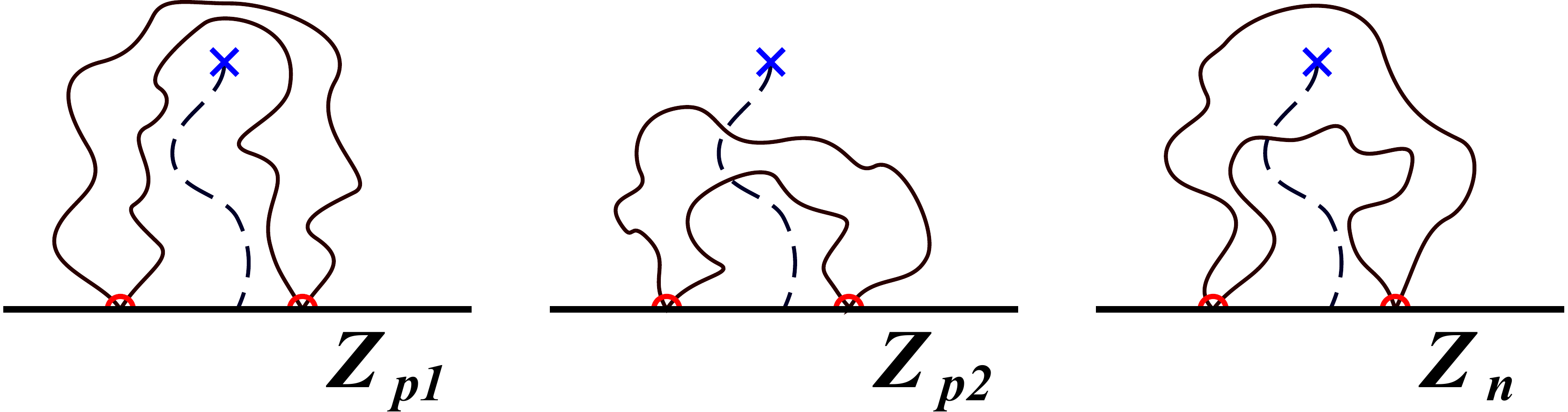} 
   \caption{The two different relevant contributions of the double path (Red circles) process with respect to the defect point (Blue Crosses).}
   \label{BCFT2}
\end{figure}

Now reexamine the correlation function
\be
\langle \phi_{2,1}(z,\bar z)\phi_{1,3}(x_1) \phi_{1,3}(x_2) \rangle\; ,
\ee
at $n=0$.  Conditioning the system on the existence of a double SLE$_{8/3}$ we have partition function
\be
Z=Z_{p1}+Z_n+Z_{p2} = \langle \phi_{1,3}(x_1) \phi_{1,3}(x_2) \rangle=(x_2-x_1)^{-4}
\ee
and the total mass of these configurations may be decomposed into three parts as in Fig.\ \ref{BCFT2}.  

Our previous result (\ref{2xSLEEq}) tells us that in the strip
\be
\frac{\langle \phi_{2,1}(s,\bar s)\phi_{1,3}(-\infty) \phi_{1,3}(\infty) \rangle_{\mathbb{S}}}{\langle\phi_{1,3}(-\infty) \phi_{1,3}(\infty) \rangle_{\mathbb{S}} }  =\frac{Z_{p1}-Z_n+Z_{p2} }{Z}= \left(1-\frac{8}{5}  \sin^2(v) \right)\; ,
\ee
from which we determine the probability that the twist operator lies between the two SLEs
\be \label{doubleSLEsingletwist}
P_n(u)=\frac{Z_n}{Z}=\frac{4}{5}  \sin^2(v)\; .
\ee
This matches the extension of Schramm's formula to double SLEs derived in \cite{GamsaCardy05} for $\k =8/3$.  We are unable to disentangle the contributions from $Z_{p1}$ and $Z_{p2}$ using twist operators because of their parity symmetry.

In this section and the last we've reported two twist operator results that were previously derived using a zero-weight indicator operator.  Based on the physical interpretation of the twist operator we expect that it should mimic the indicator operator in these cases, so the agreement indicates that the assertion in \cite{GamsaCardy06} that the second level null-state should be set to zero is consistent.  In the next section we will use this null-state as an extra condition to determine our correlation function. Furthermore, with only one path in the configuration there will be no parity issues and the twist operator should play the role of the indicator operator perfectly.

%%%%%%%%%%%%%%%%%%%%
%                  NEW RESULT !!!                  %
%%%%%%%%%%%%%%%%%%%%
\section{A New SLE$_{8/3}$ Result \label{NewResult}}

\begin{figure}[htbp] %  figure placement: here, top, bottom, or page
   \centering
   \includegraphics[width=5in]{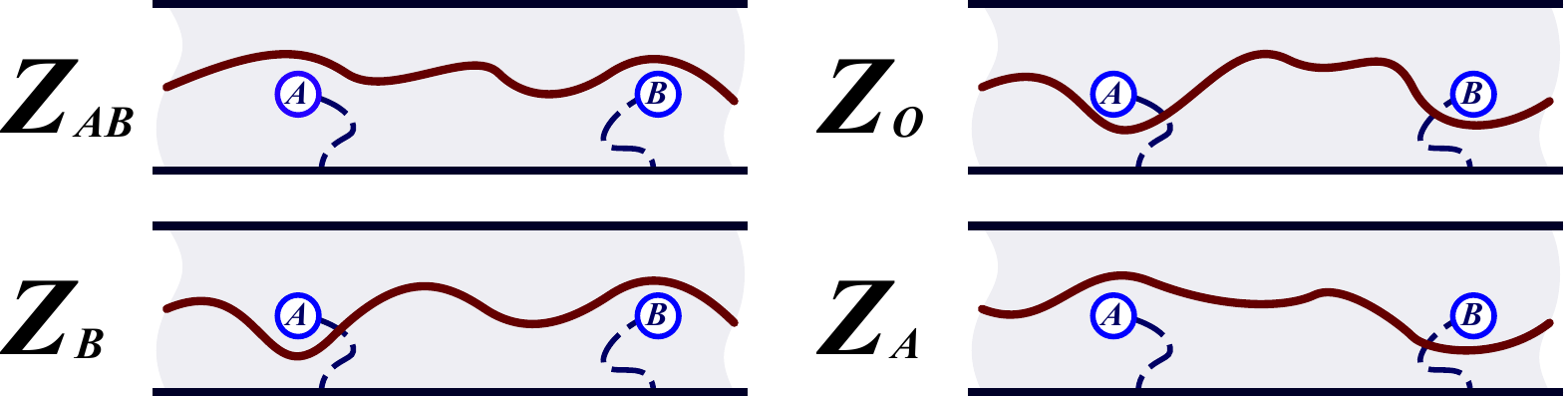} 
   \caption{Schematic representation of the four possible winding states of the SLE with respect to the positions $z_A$ and $z_B$.}
   \label{Schematic1}
\end{figure}

In the last section we showed that at $\k = 8/3$ the twist operator can be used in place of Schramm's zero weight indicator operator do derive various probablities. In this section we describe how twist operators can be used to calculate a new SLE$_{8/3}$ result.  We decompose the total weight of all SLE paths in the infinite strip according to the relative winding around two points, labeled by $A$ and $B$ in Fig.\ \ref{Schematic1}.
We denote these weights by $Z_i$, where the index $i$ labels which of $A$ and/or $B$ are adjacent to the bottom edge of the strip.

We can write down down four possible correlation functions based on whether or not we place twists operator at $z_A$ and/or $z_B$.  We fix all signs so that the twists parity is measured with respect to the bottom edge.  The four correlation functions and their expressions in terms of our decomposition are
\bea \label{0Twist}
\frac{Z_{AB}+Z_A+Z_B+Z_O}{Z}&=&\frac{\langle \phi_{1,2}(-\infty) \phi_{1,2}(\infty) \rangle_{\mathbb{S}}}{\langle \phi_{1,2}(-\infty) \phi_{1,2}(\infty) \rangle_{\mathbb{S}}}\; =\;1\\ \label{ATwist}
\frac{Z_{AB}+Z_A-Z_B-Z_O}{Z}&=&\frac{\langle \phi_{2,1}(s_A,\bar s_A) \phi_{1,2}(-\infty) \phi_{1,2}(\infty) \rangle_{\mathbb{S}}}{\langle \phi_{1,2}(-\infty) \phi_{1,2}(\infty) \rangle_{\mathbb{S}}}\; =\; \cos(v_A)\\ \label{BTwist}
\frac{Z_{AB}-Z_A+Z_B-Z_O}{Z}&=&\frac{\langle \phi_{2,1}(s_B,\bar s_B) \phi_{1,2}(-\infty) \phi_{1,2}(\infty) \rangle_{\mathbb{S}}}{\langle \phi_{1,2}(-\infty) \phi_{1,2}(\infty) \rangle_{\mathbb{S}}}\; =\; \cos(v_B)\\  \label{ABTwist}
\frac{Z_{AB}-Z_A-Z_B+Z_O}{Z}&=&\frac{\langle \phi_{2,1}(s_A,\bar s_A) \phi_{2,1}(s_B,\bar s_B) \phi_{1,2}(-\infty) \phi_{1,2}(\infty) \rangle_{\mathbb{S}}}{\langle \phi_{1,2}(-\infty) \phi_{1,2}(\infty) \rangle_{\mathbb{S}}} \\ \nonumber
&=&\cos(v_A)\cos(v_B)+\sin(v_A)\sin(v_B) \left(1-\sigma\,{}_2F_1\left(1,\frac{4}{3};\frac{5}{3};1-\sigma \right)\right)\; ,
\eea
with the parameter $\sigma$ defined as
\be
\sigma =\frac{\cosh(u_B-u_A)-\cos(v_A-v_B)}{\cosh(u_B-u_A)-\cos(v_A+v_B)}
\ee

Of these equations the first three are either trivial or equivalent to (\ref{SLEsingletwist}). However, Eq.\ (\ref{ABTwist}) is new, and can be shown to be the unique  function which obeys both the boundary $\phi_{1,2}$ null-state and bulk $\phi_{2,1}$ null-state differential equations, while satisfying the limiting condition
\be \label{limcond}
\lim_{u_2-u_1 \to \infty} \frac{\langle \phi_{2,1}(s_A,\bar s_A) \phi_{2,1}(s_B,\bar s_B) \phi_{1,2}(-\infty) \phi_{1,2}(\infty) \rangle_{\mathbb{S}}}{\langle \phi_{1,2}(-\infty) \phi_{1,2}(\infty) \rangle_{\mathbb{S}}} = \cos(v_A)\cos(v_B)\; .
\ee
We outline the derivation of this new quantity beginning with the form implied by conformal symmetry in the upper-half plane:
\be
\langle \phi_{2,1}(z_A,\bar z_A) \phi_{2,1}(z_B,\bar z_B) \phi_{1,2}(x_1) \phi_{1,2}(x_2) \rangle = (x_2-x_1)^{-5/4}\l\, \m\, F\left( \l,\m, \n \right)\; ,
\ee
with variables chosen as functions of cross-ratios that simplify the dependance on the strip variables,
\bea \label{lamb}
\l\; =\; \cos(v_A) &=& \frac{(z_A-x_1)(x_2-\bar z_A)+(\bar z_A-x_1)(x_2-z_A)}{2\, |z_A-x_1|\, |x_2-z_A|}\; ,\\ \label{mu}
\m\; =\; \cos(v_B) &=& \frac{(z_B-x_1)(x_2-\bar z_B)+(\bar z_B-x_1)(x_2-z_B)}{2\, |z_B-x_1|\, |x_2-z_B|}\qquad \mathrm{and,}\\
\n\; =\; \cosh(u_B-u_A) &=& \frac{|z_A-x_1|^2|x_2-z_B|^2+|z_B-x_1|^2|x_2-z_A|^2}{2 |z_B-x_1|\,|x_2-z_A|\,|z_A-x_1|\,|x_2-z_B|}\; .
\eea
Each operator in the correlation function has a second order null state, leading to six partial differential equations for $F(\l,\m,\n)$. %and then taking the limit 
%\be
%\{x_1, x_2, z_A, \bar z_A,  z_B, \bar z_B\} \to \{0,\, \infty,\, e^{u_A+i v_A},\, e^{u_A-i v_A},\, e^{u_B+i v_B},\, e^{u_B-i v_B}\}\; ,
%\ee
%and simplifying the results.
These equations are cumbersome and we do not record them here.  However, they can be rearranged to yield
\be \label{bigDE1}
0=\n\, (\l^2-\m^2)\, \partial_{\n} F(\l,\m,\n)=\l\, (1-\l^2) \partial_{\l} F(\l,\m,\n)-\m\, (1-\m^2) \partial_{\m} F(\l,\m,\n)\; .
\ee
If we make the change of variables
\bea
\l\, \m &=& \rho\; =\; \cos(v_A)\cos(v_B)\\
\frac{\n-\l\, \m -\sqrt{(1-\l^2)(1-\m^2)}}{\n-\l\, \m +\sqrt{(1-\l^2)(1-\m^2)}} &=& \sigma\; =\; \frac{\cosh(u_B-u_A)-\cos(v_B-v_A)}{\cosh(u_B-u_A)-\cos(v_B+v_A)}\\
\frac{\sqrt{(1-\l^2)(1-\m^2)}}{\l\, \m} &=& \tau\; =\; \tan(v_A) \tan(v_B)\; ,
\eea
then (\ref{bigDE1}) implies $0=\partial_{\rho} F(\rho,\sigma,\tau)$ indicating a reduction in the effective number of variables
\be
F(\l,\m,\n)=H(\sigma,\tau)\; .
\ee
In terms of this new function the original differential equations are equivalent to
\bea
0&=&\partial_\tau{}^2 H(\sigma,\tau)\qquad \mathrm{and}\\
0&=&(1-\sigma)\tau H(\sigma,\tau)-\tau(1+\tau+\sigma-\sigma \tau)\partial_\tau H(\sigma, \tau)+3\sigma(1-\sigma)\partial_\sigma H(\sigma,\tau)\; ,
\eea
with two dimensional solution space
\be \label{Hexp}
H(\sigma, \tau)=c_1\left(1+\tau \left(1-\sigma\,{}_2F_1\left(1,\frac{4}{3};\frac{5}{3}\bigg|1-\sigma \right)\right)\right)+c_2\, \tau \left( \frac{\sigma}{(1-\sigma)^2}\right)^{1/3}\; .
\ee
To pick out the physical solution we enforce the limit (\ref{limcond}), or equivalently
\be
\lim_{\sigma \to 1} H(\sigma,\tau)= 1\; .
\ee
Since the second term diverges in this limit we require that $c_{2}=0$ and (\ref{ABTwist}) is the result of setting $c_{1}=1$ and restoring the other factors to the correlation function.

\begin{figure}[htbp] %  figure placement: here, top, bottom, or page
   \centering
   \includegraphics[width=3in]{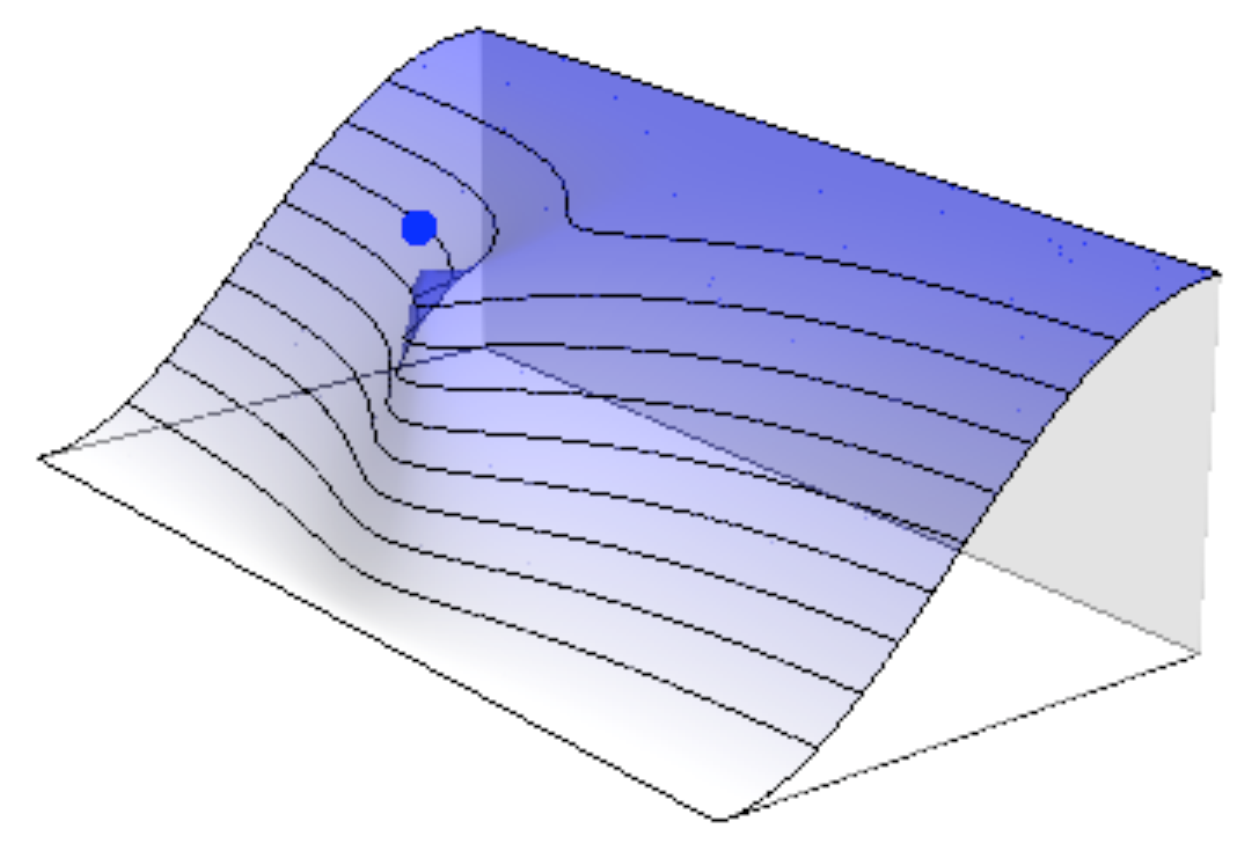} 
   \caption{This is the probabilty that the SLE path travels below (towards the near edge of this plot) the two twist operators as a function of position of $z_B$, given that $z_A$ is fixed under the blue dot.}
   \label{ProbDens}
\end{figure}

With (\ref{0Twist}--\ref{ABTwist}) we can construct the probabilities of the winding configurations in Fig.\ \ref{Schematic1}, $P_i = Z_i/Z$,
\bea \label{WABStrip}
P_{AB}&=&\frac{1}{4}\left( (1+\cos(v_A))(1+\cos(v_B))+\sin(v_A)\sin(v_B)\left(1-\sigma\,{}_2F_1\left(1,\frac{4}{3};\frac{5}{3}\bigg|1-\sigma \right)\right)\right)\; ,\\ \label{WAStrip}
P_{A} &=& \frac{1}{4}\left( (1+\cos(v_A))(1-\cos(v_B))-\sin(v_A)\sin(v_B) \left(1-\sigma\,{}_2F_1\left(1,\frac{4}{3};\frac{5}{3}\bigg|1-\sigma \right)\right)\right)\; ,\\ \label{WBStrip}
P_{B} &=& \frac{1}{4}\left( (1-\cos(v_A))(1+\cos(v_B))-\sin(v_A)\sin(v_B)\left(1-\sigma\,{}_2F_1\left(1,\frac{4}{3};\frac{5}{3}\bigg|1-\sigma \right)\right)\right)\; ,\\ \label{W0Strip}
P_{O} &=& \frac{1}{4}\left( (1-\cos(v_A))(1-\cos(v_B))+\sin(v_A)\sin(v_B) \left(1-\sigma\,{}_2F_1\left(1,\frac{4}{3};\frac{5}{3}\bigg|1-\sigma \right)\right)\right)\; .
\eea

These probabilities are conformally invariant and apply to the upper-half plane as well as in the strip.  If the SLE in the upper half plan runs from the origin to infinity then we only need to know that $v_{A(B)}$ is the argument of $z_{A(B)}$ and that in the half-plane
\be \label{sigma}
\sigma = \frac{(z_B-z_A)(\bar z_B-\bar z_A)}{(z_B-\bar z_A)(\bar z_B-z_A)} = \frac{r_B{}^2-2 r_A r_B\cos(v_B-v_A)+r_A{}^2}{r_B{}^2-2 r_A r_B\cos(v_B+v_A)+r_A{}^2}\; .
\ee

%%%%%%%%%%%%%%%%%%%%
%                          LCFT                              %
%%%%%%%%%%%%%%%%%%%%
\section{Logarithmic Conformal Field Theory \label{LCFT}}

In the last section we saw that the collection of chiral operators $\langle \phi_{2,1} \phi_{2,1} \phi_{2,1}\phi_{2,1}\phi_{1,2}\phi_{1,2}  \rangle$ allows two linearly independent  solutions.   This correlation function is non-trivial and consistently satisfies all of the null-state conditions of both $\phi_{1,2}$ and $\phi_{2,1}$ operators.  To see why this is interesting we recall the discussion of \cite{MathieuRidout07} involving the logarithmic structure of the CFT generated by $\phi_{2,1}$ and raising the question of whether this CFT is compatible with that generated by $\phi_{1,2}$.  Specifically we note that in our physical interpretation of this correlation function, the two different species of operators appear as either bulk or boundary operators, but never as both.  We speculate that this leads to a resolution of the co-existence issue and an explanation of the disagreement between \cite{GurarieLudwig00}  and \cite{MathieuRidout07}  predicting which logarithmic partner appears in the self avoiding walk and percolation models.

In what follows note that our Kac indices are in reverse order to those in \cite{MathieuRidout07} because we use the SLE$_{8/3}$ convention while they work with the dual SLE$_{6}$ convention.

We begin with the sector of our CFT generated by the Verma module $\mathfrak{M}_{2,1}$ of which $\phi_{2,1}$ with weight $h_{2,1}=0$ is the highest weight state.  Fusing these modules together leads to a variety of staggered logarithmic modules, 
\bea
\mathfrak{M}_{2,1} \times \mathfrak{M}_{2,1} &=& \mathfrak{M}_{1,1}+\mathfrak{M}_{3,1}\; ,\\
\mathfrak{M}_{2,1} \times \mathfrak{M}_{2,1} \times \mathfrak{M}_{2,1} &=& \mathfrak{M}_{2,1}+\mathfrak{I}_{4,1}\; ,\\
\mathfrak{M}_{2,1} \times \mathfrak{M}_{2,1} \times \mathfrak{M}_{2,1} \times \mathfrak{M}_{2,1} &=& \mathfrak{M}_{1,1}+3\, \mathfrak{M}_{3,1}+\mathfrak{I}_{5,1}\; ,
\eea
but most notably we are interested in the staggered module $\mathfrak{I}_{5,1}$ which contains $\phi_{5,1}$, a candidate for the logarithmic partner to the stress energy tensor.

The other potential logarithmic partner $\phi_{1,3}$, occurs in the sector generated by $\mathfrak{M}_{1,2}$ with highest weight $h_{1,2}=5/8$,
\bea
\mathfrak{M}_{1,2} \times \mathfrak{M}_{1,2} &=& \mathfrak{I}_{1,3}\; .
\eea

Both of these logarithmic partners behave in a similar way under the action of the Virasoro generators
\be
L_0|\phi\rangle = 2 |\phi\rangle +|T\rangle\; , \qquad L_1|\phi\rangle = 0\; , \qquad L_2|\phi\rangle = \b |1\rangle,
\ee
but the two point function $\langle \phi_{1,3} \phi_{5,1} \rangle$ is undefined because the two constants $\b_{1,3}=5/6$ and $\b_{5,1}=-5/8$ are not the same \cite{GurarieLudwig00}.  This has lead to speculation that $\mathfrak{M}_{2,1}$ and $\mathfrak{M}_{1,2}$ cannot both be included in a well defined theory, because otherwise the closure of the theory would necessarily include undefined objects.

Now that we have an example of a physical correlation function built from these two modules we can examine the underlying LCFT structure.  We first separate our operators by species and examine the lowest order terms of the operator product expansions (OPEs) for the identity sector:
\be
\phi_{1,2}(u)|\phi_{1,2}\rangle= C_{1,3} u^{-5/4} \left(|1\rangle+\frac{3}{4}u^2\left( |\phi_{1,3}\rangle+\log(x) |T\rangle \right) \right)+ C_T u^{3/4} |T\rangle +\mathcal{O}(u^{7/4})\; ,
\ee
with two degrees of freedom related to the stress tensor and it's partner in $\mathfrak{I}_{1,3}$, and
\bea
\phi_{2,1}(x)\phi_{2,1}(y)\phi_{2,1}(z)|\phi_{2,1}\rangle&=& \left( C_1+C_{5,1}\left(\frac{(x-y)z}{y(x-z)} \right)^{1/3}{}_2F_1\left(\frac{1}{3},\frac{2}{3},\frac{4}{3} ;\frac{(x-y)z}{y(x-z)}\right) \right) |1\rangle\\ \nonumber
&&\phantom{C_{5,1}\Bigg(}-\frac{8 C_{5,1}}{15}(x y z (x-y)(y-z)(x-z))^{1/3}\left(|\phi_{5,1}\rangle+\log(x) |T\rangle \right)\\ \nonumber
&&+ F_T\left(\frac{y}{x},\frac{z}{x}\right) x^2 |T\rangle +\mathcal{O}(x^3)\; ,
\eea
with three degrees of freedom related to the identity in $\mathfrak{M}_{1,1}$ and the stress tensor and it's partner in $\mathfrak{I}_{5,1}$% (we've neglected three copies of $\mathfrak{M}_{3,1}$ that cannot couple to $\mathfrak{I}_{1,3}$ because of their incompatible weights $h_{3,1}=1/3$ and $h_{1,3}=2$)
.  The OPEs are determined from the action of the Virasoro generators on an OPE Ansatz including the appropriate logarithmic partner field and further constrained using the second order null-state.  Of the five parameters, $ C_1, C_{3,1}, C_{5,1}$, and $C_T$ are constants that depend on the fusion channel of the ensemble.  The fusion channel determines $F_T(y/x,z/x)$ as well via the differential equation
\bea \label{C2DE}
C_1&=&\left(5+\left( \frac{(1-u)(2+11u)}{2u}\partial_u+\frac{(1-v)(2+11v)}{2v}\partial_v \right)+\frac{3}{2}\left( (1-u)\partial_u+(1-v)\partial_v \right)^2 \right) F_T(u,v)\\ \nonumber
&&-C_{5,1}\left( 4 \frac{2u+2v+u v}{15 u v} \left(u (1-v)v(u-v)(1-u)\right)^{1/3}+\left(\frac{v(1-u)}{u(1-v)}\right)^{1/3}{}_2F_1\left(\frac{1}{3}, \frac{2}{3}; \frac{4}{3}; \frac{v(1-u)}{u(1-v)}\right)\right)\; .
\eea

Now, the two conformal blocks included in (\ref{Hexp}) are defined about $\sigma=1$, which we recall corresponds to the limit of large separation along the length of the strip.  The crossing symmetry relation
\bea \label{XingSym}
\left[ \rho+\rho\,\tau \left(1-\sigma\,{}_2F_1\left(1,\frac{4}{3};\frac{5}{3}\bigg|1-\sigma \right)\right)\right]_{B_1}&=&\left[ \rho+\rho\,\tau \left(1+\sigma\,{}_2F_1\left(1,\frac{4}{3};\frac{5}{3}\bigg|\sigma \right)\right)\right]_{B_3}\\ \nonumber
&&-\left[\frac{\Gamma(5/3)\Gamma(2/3)\, \rho\,\tau\,\sigma^{1/3}}{\Gamma(4/3)(1-\sigma)^{2/3}}\right]_{B_2} ,
\eea
allows us to identify the blocks about $\sigma=0$, the limit when the twist operators meet in the middle of the strip.  We name the blocks according to the bracket labels in (\ref{XingSym}), so that $B_1$ and $B_2$ are the $\sigma=1$ blocks and $B_2$ and $B_3$ are the $\sigma=0$ blocks.

We can determine the fusion channels that contribute to the observed conformal blocks by taking
\be
\frac{\langle \phi_{1,2}(\xi+\delta) \phi_{1,2}(\xi)  \phi_{2,1}(x) \phi_{2,1}(y) \phi_{2,1}(z) \phi_{2,1}(0) \rangle}{ \langle  \phi_{1,2}(\xi+\delta) \phi_{1,2}(\xi)  \rangle} = B_j
\ee
with $\xi \gg \delta, x$ and $x>y>z>0$ and performing a series expansion in $1/\xi$. In terms of these variables, and to second leading order in $1/\xi$, the parameters of the last section are
\bea
\rho &=& 1+\frac{\delta^2}{8 \xi^4}\left((y-z)^2+x^2\right) +\mathcal{O}(\xi^{-5})\\
\rho\, \tau &=&-\frac{\delta^2}{4 \xi^4}\left(x(y-z)\right) +\mathcal{O}(\xi^{-5})\\
\sigma&=&\frac{z(x-y)}{y(x-z)}\; .
\eea
The conformal blocks are
\bea
B_1&=&\left(1+\frac{\delta^2}{8 \xi^4}\left((x-y+z)^2+2 \frac{x(x-y)(y-z)z}{(x-z)y}\,{}_2F_1\left(1,\frac{4}{3};\frac{5}{3}\bigg|\frac{x(y-z)}{y(x-z)} \right)\right)\right)+ \mathcal{O}\left(\xi^{-5} \right)\\
B_2&=&-\frac{\Gamma(5/3)\Gamma(2/3)}{\Gamma(4/3)}\frac{\delta^2}{4 \xi^4}\left(x(x-z)y(x-y)(y-z)z\right)^{1/3} + \mathcal{O}\left(\xi^{-5} \right)\\
B_3&=&\left(1+\frac{\delta^2}{8 \xi^4}\left((x-y+z)^2-2 \frac{x(x-y)(y-z)z}{(x-z)y}\,{}_2F_1\left(1,\frac{4}{3};\frac{5}{3}\bigg|\frac{z(x-y)}{y(x-z)} \right)\right)\right)+ \mathcal{O}\left(\xi^{-5} \right)\; .
\eea
These expressions let us determine the corresponding fusion channels from the OPEs and two point functions
\bea
\langle \phi(\xi) T(0) \rangle=\langle T(\xi) \phi(0) \rangle=\b/\xi^4\quad \mathrm{and} \quad \langle T(\xi) T(0) \rangle=0\;   .
\eea

First, let's take block $B_1$: it has an order $\xi^0$ term, which requires that $C_{1,3} \neq 0$ in the $\left( \phi_{1,2} \right)^2$ fusion.  Furthermore, the leading term is constant so that $C_{5,1}=0$ and thus $C_1 \neq 0$.  As a double check we can successfully verify that the form of $F_T(u,v)$ implied by the order $\xi^{-4}$ term satisfies (\ref{C2DE}) with $C_1 \neq 0$ and $C_{5,1}=0$.  The final coefficient, $C_T$ cannot be determined because of the lack of a logarithmic partner for this stress tensor, but it is clear that the operator that plays the role of the logarithmic partner is $\phi_{1,3}$ and that the $\phi_{5,1}$ channel is absent from the $\left( \phi_{2,1} \right)^4$ fusion.  This reinforces the assertion, made in \cite{MathieuRidout07}, that the boundary LCFT describing self avoiding loops should contain the logarithmic partner with parameter $\b_{1,3}=5/6$.

The block $B_3$ has a structure very similar to $B_1$, but with the notable exception that the the blocks have different forms for $F_T(u,v)$ though, of course, both satisfy (\ref{C2DE}).

In comparison the block $B_2$ has leading term of order $\xi^{-4}$, so either $C_{1,3}=0$ or $C_1 = C_{5,1} = 0$.  One of the fusions includes the logarithmic partner field and one only the stress tensor.  It is not possible to determine which logarithmic partner is included or omitted using only the form of the conformal block because while the coefficient of $| \phi_{5,1} \rangle$ is of the correct form, the same form also solves (\ref{C2DE}) with $C_1=C_{5,1}=0$.  However, in  order to ensure consistency under crossing symmetry this block must have the same logarithmic partner as $B_1$ and $B_3$.  Thus $C_{1,3}\ne0$ and $C_1 = C_{5,1} = 0$ for this block.

\begin{figure}[htbp] %  figure placement: here, top, bottom, or page
   \centering
   \includegraphics[width=6in]{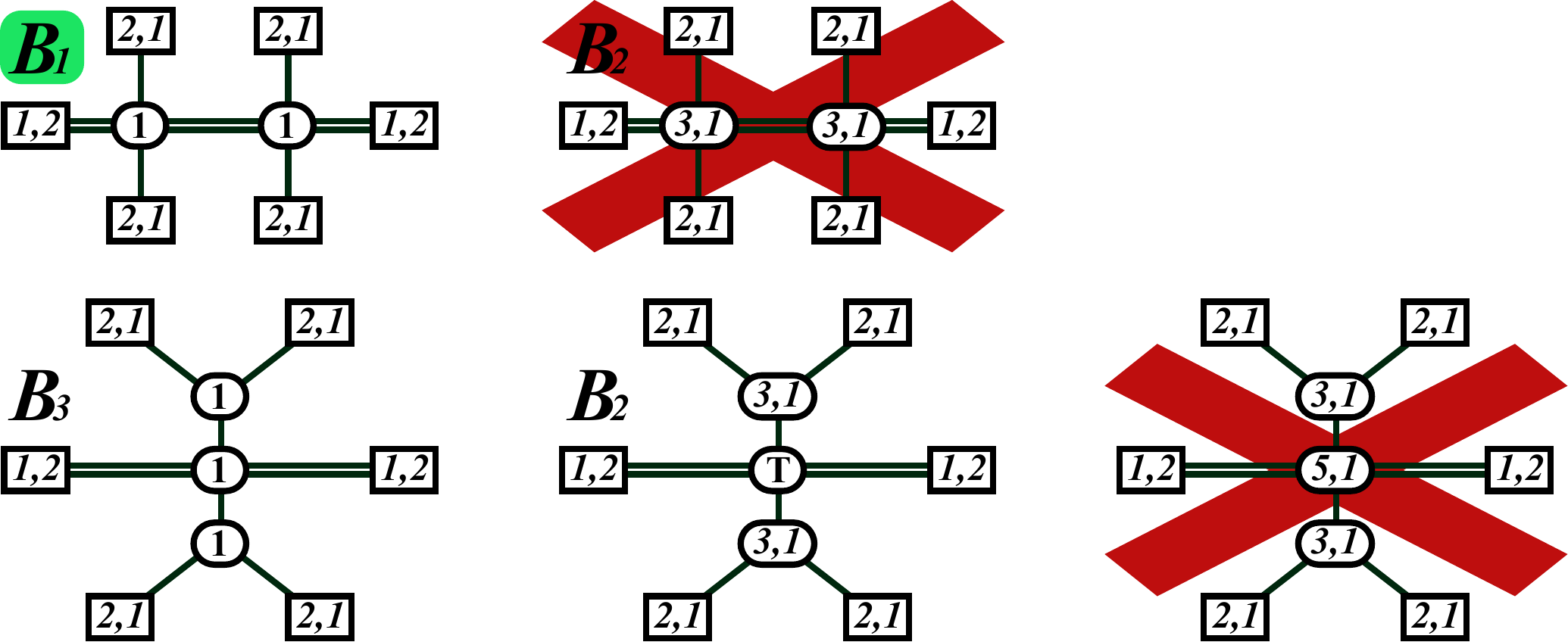} 
   \caption{These are the conceivable conformal blocks of $\langle \phi_{1,2}(x_1)\phi_{1,2}(x_2)\phi_{2,1}(z_A, \bar z_A)\phi_{2,1}(z_B, \bar z_B) \rangle$, \emph{i.e.} single valued, with (potentially) compatible Kac labels.  The square boxes denote chiral operators and oval boxed denote fusion products.  The double line links the boundary operators, including bulk-boundary fusion products, while the operators above (below) the double line represent (anti-)holomorphic parts of the bulk operators.  We do not resolve fusions between boundary operators because these are either unique or logarithmically ambiguous; in the case of logarithmic ambiguity we mark the diagram with a bold red ``X''.  The blocks are labeled as in the text and the physical block is highlighted in green.}
   \label{SAWCB}
\end{figure}

The leading weight of the two $\sigma \to 0$ blocks are $h_{1,1}=0$ and $h_{3,1}=1/3$, and the physical solution is equivalent to the identity block minus the $\phi_{3,1}$ block.   This agrees favorably with the calculated correlation function of twist operators and an anchored loop where we found the one physical solution, $|\mathcal{F}_{1,1}(x)|^2-|\mathcal{F}_{3,1}(x)|^2$, also the identity block minus the $\phi_{3,1}$ block.

As we noted previously, the results of the last section confirm that the logarithmic partner to the boundary stress tensor is $\phi_{1,3}$ with associated parameter $\beta_{1,3}=5/6$.  This is in agreement with the prediction of Mathieu and Ridout in \cite{MathieuRidout07}, which seemed to contradict the earlier assertion of Gurarie and Ludwig in \cite{GurarieLudwig00} that the logarithmic partner to the stress tensor for the self avoiding walk is $\phi_{5,1}$.  However, Gurarie and Ludwig made their assertion based on correlation functions of bulk operators and in fact the results of \cite{GamsaCardy06} verify the assertion of Gurarie and Ludwig.  This is because the conformal blocks used to assemble the correlation function in \cite{GamsaCardy06} are equivalent to those appearing in the derivation for Cardy's and Watts' formulae, and it was argued in \cite{MathieuRidout07} that the appearance of these blocks requires the inclusion of $\phi_{5,1}$ in the corresponding CFT.  

Thus it seems that we desire a logarithmic BCFT with two logarithmic partners of $T(z)$: $\phi_{5,1}$ in the bulk, and $\phi_{1,3}$ on the boundary.    The only way this is possible is if the bulk operators fuse to the boundary exclusively via the identity module.  Specifically this is a statement about the bulk-boundary OPE for the class of conformal boundary conditions that support SLE type boundary conditions with $\phi_{1,2}$ acting as a boundary condition changing operator, it is possible that other boundary conditions exist which do not support $\phi_{1,2}$ and which allow a fuller bulk-boundary fusion content.

The physical solution given by (\ref{ABTwist}) is represented by the block $B_1$, as indicated in Fig.\ \ref{SAWCB},  and can be written as the difference between blocks $B_3$ and $B_2$.  The blocks marked with a red ``X'' require bulk-boundary fusion that is inconsistent with our proposed boundary conditions.  This means the two blocks $B_2$ and $B_3$ cannot exist independently, because under crossing symmetry this would require inclusion of the right most block in the upper row.  But excluding $B_2$ and $B_3$ seems natural because they are both divergent in the limit where $u_1-u_2 \to \infty$ and therefore unphysical.

%%%%%%%%%%%%%%%%%%%%
%                PERC. SUBSECT.                 %
%%%%%%%%%%%%%%%%%%%%
\subsection*{The implications of this result on $\k = 6$ and critical percolation}

In contrast to the case already discussed above, we may consider the result of insisting that fusions occur via the $\phi_{5,1}$ channel.  In this case the $B_1$ block would be excluded and the $B_2$ block would be the only solution, which is self consistent under crossing symmetry, as we pointed out previously.  Whereas this new case is speculated to correspond to critical percolation, we point out that this chiral correlation function does indeed have a physical interpretation in this model.  The Kac indices in this section are reversed from those in the last because we are working with an SLE$_6$ formalism.  Thus, in the percolation model the four zero weight operators correspond to SLE operators $\phi_{1,2}$, and the first element of the correlation function is a set of four paths attached to the boundary. 

As for the two $5/8-$weight operators, note that the identification of $\phi_{2,1}$ as a twist operator doesn't survive as we enter into the dense phase.  In the dense phase each bond is occupied and there is no difference in loop parity between different configurations, so $\phi_{2,1}$ belongs to the identity sector as the leading bulk energy density operator, a role it takes over from $\phi_{3,1}$ in the dilute phase.  But consider for a moment, that in the $Q-$state Potts model (of which percolation is the $Q \to1$ limit) the energy density operator can be identified as the fusion product of two different species of spin operators.  For these two spins to be distinct the outer edges of two clusters must both pass through the point, implying that a total of four legs emanate from the energy density operator.  At $Q=1$ the weights of the loop configurations are independent of our spin labels and the energy density operator becomes equivalent to the $4-$leg operator which was identified as $\phi_{0,2}$ in \cite{SaleurDuplantier87}.  

Therefore the configurations in the boundary percolation model described by the six-point function $\langle  \phi_{1,2}(x_1)\phi_{1,2}(x_2)\phi_{1,2}(x_3)\phi_{1,2}(x_4)\phi_{2,1}(z, \bar z) \rangle $ are those with four SLE type paths flowing towards a single bulk point, a multiple radial SLE$_6$.  As we bring the bulk $4-$leg operator to the boundary we naturally expect it's bulk boundary fusion to have a boundary $4-$leg operator leading term corresponding to the boundary operator with weight $h_{1,5}=2$.  But $\phi_{1,5}$ cannot appear in the $\phi_{2,1} \times \phi_{2,1}=\mathfrak{I}_{3,1}$ fusion so the leading operator must be the stress tensor.  
\begin{figure}[htbp] %  figure placement: here, top, bottom, or page
   \centering
   \includegraphics[width=5in]{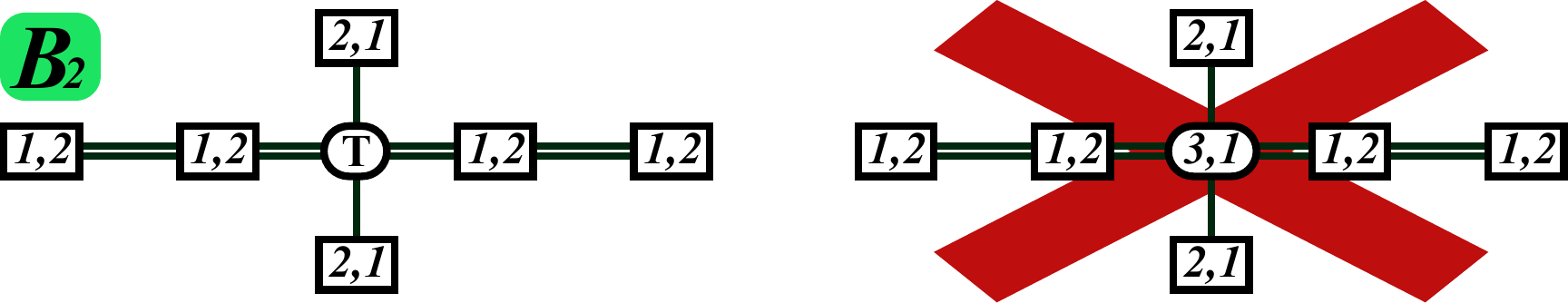} 
   \caption{Using the same conventions as Fig.\ \ref{SAWCB} we illustrate the potential conformal blocks of our percolation correlation function.}
   \label{PercCB}
\end{figure}
This is indeed the fusion channel of the $B_2$ block, which leads to the correlation function
\be
\frac{\mathrm{Im}(z)^{3/4} (x_4-x_3)^{1/3}(x_4-x_2)^{1/3}(x_4-x_1)^{1/3}(x_3-x_2)^{1/3}(x_3-x_1)^{1/3}(x_2-x_1)^{1/3}}{|z-x_4|\,|z-x_3|\,|z-x_2|\,|z-x_1|}\; .
\ee

Again we see that the allowed block is formed by insisting the bulk CFT fuses to the boundary CFT exclusively via the identity block, with a different conformal block implied because of the interchange of bulk and boundary sub-theories between the two models.

We end by suggesting that this construction may apply to a wider variety of extended minimal model BCFTs.  It is perhaps natural to assume that two different logarithmic partners may reside in the bulk and on the boundary and that the consistency of the theory may be based on restricting the allowed bulk-boundary fusions.  But then the allowed fusion products should include the entire minimal model Kac table, since we expect the extended CFT to include the minimal model as a sub-sector.  This is essentially what we've done above, since the $c=0$ the minimal model is simply the identity operator.

%%%%%%%%%%%%%%%%%%%%
%                  CONCLUSIONS                  %
%%%%%%%%%%%%%%%%%%%%
\section{Conclusions \label{Conc}}

In this paper we have assembled a variety of  correlation functions with in the $O(n)$  containing twist operators.  The correlation functions included correspond to: a loop anchored at two points with a pair of twist operators, an SLE with a pair of twist operators, and a double SLE with a pair of twist operators.  We then take these correlation functions in the $n \to 0$ limit where the $O(n)$ model corresponds to self avoiding loops or dilute polymers.  We verify that in this limit the twist operator can play the role of zero weight indicator operator introduced by Schramm by confirming that the limiting forms of the SLE and twist operator correlation functions are equivalent to known results using the indicator operator.  

We then use the extra constraints imposed by the twist operator to solve the correlation function of a new SLE type result with two identified bulk points.  This result determines the winding distribution of an SLE$_{8/3}$ with respect to two marked boundary points.

In addition to determining a novel result for the self avoiding walk, the calculation of the conformal blocks for the chiral $6-$point function $\langle \phi_{2,1} \phi_{2,1} \phi_{2,1}\phi_{2,1}\phi_{1,2}\phi_{1,2}  \rangle$ allow us to discuss a relevant problem in the logarithmic CFT with $c=0$.  The problem arises from the inability of both $\mathfrak{M}_{1,2}$ and $\mathfrak{M}_{2,1}$ to exist in the same chiral theory, in spite of the fact that both of these modules correspond to meaningful observable which can coexist in the self avoiding loop and percolation models, both of which correspond to $c=0$.  In this paper we conjecture that the resolution to this apparent contradiction is that the two incompatible modules must belong to different sectors whether bulk or boundary, and that the bulk-boundary fusions must be through the identity channel common to both.  This is consistent with the conformal blocks found in the text and their interpretations as physical quantities.  It is our intention to further catalogue the roles played by the logarithmic operators in both the CFTs of boundary percolation and the self avoiding loop in a forthcoming paper.

We further conjecture that this coexistence of incompatible modules on the bulk and boundary may also apply to other extended minimal models such as the $O(1)$ model with $c=1/2$ under the condition that the bulk-boundary fusions occur via the minimal model subsector.

\section{Acknowledgments}
This work was supported by EPSRC Grant No.\ EP/D070643/1.

\bibliography{TwistOn}
\end{document}